\documentclass[12pt,letterpaper]{article}        
\usepackage{jheppub}
\usepackage{epsfig}
\usepackage{amssymb,amsfonts}
\usepackage{amsmath}
\usepackage{ulem}

\preprint{MIT-CTP/4907}

\title{\large Thermal diffusivity and chaos in metals without quasiparticles}
\author[1]{Mike Blake,}

\author[2]{Richard A. Davison,}

\author[2,3]{and Subir Sachdev}

\affiliation[1]{Center for Theoretical Physics, Massachusetts Institute of Technology, Cambridge MA 02139, USA}
\affiliation[2]{Department of Physics, Harvard University, Cambridge MA 02138 USA}
\affiliation[3]{Perimeter Institute for Theoretical Physics, Waterloo, Ontario N2L 2Y5, Canada}

\emailAdd{mab90@mit.edu}
\emailAdd{rdavison@g.harvard.edu}
\emailAdd{sachdev@g.harvard.edu}

\abstract{We study the thermal diffusivity $D_T$ in models of metals without
quasiparticle excitations (`strange metals'). The many-body quantum chaos and transport
properties of such metals can be efficiently described by a holographic representation in a 
gravitational theory in an emergent curved spacetime with an additional spatial dimension.
We find that at generic infra-red fixed points $D_T$ is always related to parameters characterizing many-body quantum chaos: the butterfly velocity $v_B$, and Lyapunov time $\tau_L$ through $D_T \sim v_B^2 \tau_L$. The relationship holds independently of the charge density, periodic potential strength or magnetic field at the fixed point. The generality of this result follows from the observation that the thermal conductivity of strange metals depends only on the metric near the horizon of a black hole in the emergent spacetime, and is otherwise insensitive to the profile of any matter fields.}

\begin{document}

\maketitle

\pdfoutput=1
\pagestyle{plain} \setcounter{page}{1}
\newcounter{bean}
\baselineskip16pt
\section{Introduction} 
\paragraph{}Modern quantum materials, and in particular the high temperature superconductors, 
commonly display phases with metallic conduction of charge and heat, but without quasiparticle excitations to enable transport: these are the `strange metals'. A deeper understanding of strange
metals is an important goal of quantum condensed matter physics, as it is surely a pre-requisite
for predicting the high critical temperature of the superconductivity which emerges out
of the strange metal as the temperature is lowered. 

Theories of strange metals have used several different approaches. 
Fermions at a non-zero density possess a Fermi surface of quasiparticle excitations, and a strange
metal state can be obtained by destroying the quasiparticles via a coupling to a critical bosonic order parameter or an emergent gauge field \cite{SungSik17}. However such theories flow to strong coupling, making it difficult to develop a physical understanding of transport. In the holographic approach, 
extrapolations from dualities emerging from string theory
lead to a mapping of strange metal dynamics to a gravitational theory in an emergent curved spacetime with an extra spatial dimension \cite{Zaanen:2015oix,Hartnoll:2016apf}. Many choices for the gravitational
theories have been explored, enabling the description of different classes of strange metal states. Finally, the 
Sachdev-Ye-Kitaev (SYK) models \cite{SY92,kitaev2015talk,SS15,Gu:2016oyy,Davison:2016ngz,Balents17} employ a large $N$ limit
to obtain solvable models of strange metal transport
in the presence of disorder which self-averages. 
The SYK models have been closely connected to the holographic approaches \cite{SS10,kitaev2015talk,JMDS16b,HV16,KJ16,Jevicki16,Davison:2016ngz}, and this has considerably advanced our understanding of both approaches. In particular, there is 
a mapping from the entropy of a SYK strange metal to the
Bekenstein-Hawking entropy of black holes in the holographic emergent spacetime \cite{SS10,SS15,Davison:2016ngz}.

\paragraph{}In a separate development, remarkable connections have been pointed out between the dynamics of black holes
and the nature of quantum chaos in many-body quantum systems \cite{Shenker:2013pqa}. An
out-of-time-order correlator (OTOC) defines two chaos parameters,
the butterfly velocity $v_B$ that describes the speed at which chaos propagates, and $\tau_L$ the Lyapunov time, which controls the rate at which chaotic effects grow.
The OTOC characterizes chaos in a quantum many-body system, 
and is also connected to shock waves near the black hole horizon in the holographic dual. A precise lower-bound was established \cite{maldacena2015abound}
for the Lyapunov time $\tau_L \geq \hbar/(2 \pi k_B T)$,
where $T$ is the absolute temperature. This bound is saturated
in holographic theories with Einstein gravity, and in SYK models \cite{kitaev2015talk,Maldacena:2016hyu}. More generally, we expect states of quantum matter without
quasiparticle excitations to have $\tau_L \sim \hbar/(k_B T)$, 
while more conventional states with quasiparticles will take
longer to reach chaos with $\tau_L \sim 1/T^a$ with $a>1$ \cite{ssbook,Swingle:2016jdj,Patel:2016wdy,Chowdhury:2017jzb,Patel:2017vfp}.

\paragraph{}These advances raise the question of whether 
the chaos properties of black holes and many body systems 
are connected to transport. In particular, it was proposed in \cite{Blake:2016wvh, Blake:2016sud} that the thermoelectric diffusivities $D$ obeys
\begin{equation}
\label{eq:firsteq}
D\sim v_B^2\tau_{L}.
\end{equation}
This connection provides an appealing route to characterise the transport properties of strange metal states without quasiparticle excitations. In a normal metal, the Fermi velocity and quasiparticle mean free time are the scales relevant for transport. The relation (\ref{eq:firsteq}) suggests that the butterfly velocity and the Lyapunov time are the appropriate generalisations of these for systems without quasiparticles. Further, the lower bound on $\tau_L$ appears analogous to Mott-Ioffe-Regel bounds in quasiparticle transport \cite{Bruin2013,Hartnoll:2014lpa}.
\paragraph{}The original evidence in Ref.~\cite{Blake:2016wvh} for this proposal came from studying the charge diffusivity of systems with particle/hole symmetry, in which charge and energy diffuse independently. However, in the absence of this symmetry (e.g.~in a finite density state), the diffusive processes are coupled \cite{Hartnoll:2014lpa} and it is not clear which elements of the diffusivity matrix, if any, should be related to the chaos parameters $\tau_L$ and $v_B$. In particular, the full diffusion matrix can be sensitive to short-distance physics in the form of the momentum relaxation timescale and the thermodynamic charge susceptibility, making relations of the form (\ref{eq:firsteq}) less likely.
\paragraph{}In this paper we identify a universal piece of the diffusivity matrix that we can generically relate to the chaos exponents at infra-red fixed points. This is provided by the thermal diffusivity $D_T$ 
\begin{equation}
    D_T\equiv\frac{\kappa}{c_\rho},
    \label{thermdiffintro}
\end{equation}
where $\kappa$ is the open circuit thermal conductivity and $c_\rho=T\left(\partial s/\partial T\right)_\rho$ is the thermodynamic specific heat at fixed density, $\rho$. For a generic finite density system, this is not an eigenvalue of the diffusivity matrix but can be interpreted as the diffusivity of temperature perturbations when charge perturbations are static. However, as we justify in Appendix A, in all our fixed points we find that the Einstein relations simplify at low temperatures and the thermal diffusivity (\ref{thermdiffintro}) also corresponds to an eigenvalue of the thermoelectric diffusivity matrix.\footnote{In an isotropic magnetic field, $\kappa$ should be replaced with the magneto-thermal conductivity in equation (\ref{kappaB}) below.}

%
%
\paragraph{}This thermal diffusivity $D_T$ provides a natural candidate to relate to many body chaos. For non-zero density states, the open-circuit thermal conductivity is finite in the translationally invariant limit, and so is not sensitive to irrelevant deformations that relax momentum \cite{Mahajan:2013cja}. Moreover, the relevant thermodynamic susceptibility in this Einstein relation is the specific heat, a quantity that can be extracted from the infra-red theory. 
This intuition is reinforced by work on SYK/AdS$_2$ models  \cite{ Gu:2016oyy,Davison:2016ngz,Blake:2016jnn,Jian:2017unn,Narayan:2017qtw}  and critical Fermi surfaces \cite{Patel:2016wdy} for which connections between the thermal diffusivity and chaos have recently been observed.
\paragraph{}Here we will establish that the relationship $D_T \sim v_B^2\tau_L$ is a generic low temperature property of homogeneous holographic lattice models \cite{Donos:2014uba, Gouteraux:2014hca, Donos:2013eha, Andrade:2013gsa} that flow to infra-red fixed points. Our conclusions are robust, and are based on the simple observation that the dc conductivity formulae for these theories imply that $\kappa$ can be expressed solely in terms of the metric near the horizon.  For instance, for a $2+1$ dimensional boundary theory we find that it can always be written as 
\begin{equation}
\kappa = \frac{4 \pi f'(r_0)}{f''(r_0)}, 
\label{kappa0}
\end{equation}
where $f(r)$ is the emblackening factor of the black hole metric and $r_0$ the horizon radius. This formula holds regardless of the lattice strength, charge density or magnetic field of the boundary theory. In other words, all the dependence of $\kappa$ on these quantities is encoded in their effects on the infra-red geometry. 

\paragraph{}This property is the key to establishing our connection between the thermal diffusivity $D_T$ and the chaos parameters, which are themselves determined from the metric near the horizon. In particular for theories that flow to infra-red scaling geometries we demonstrate that this leads to 
\begin{equation}
D_T = \frac{z}{2 z-2} v_B^2 \tau_L,
\label{chaosresult}
\end{equation}
where $z$ is the dynamical critical exponent of the infra red fixed point. We emphasise that both $D_T$ and $v_B$ are typically highly non-trivial functions of the charge density, temperature, lattice strength and magnetic field. However since all that dependence is captured by the metric near the horizon, we find that they are always related by (\ref{chaosresult}) near the fixed point. Note that an exception to this general result is provided by fixed points with $z=1$, for which our expression (\ref{chaosresult}) diverges. We will discuss this special case further in section \ref{sec:diffusivity}.

\section{Thermal Conductivity in Holographic Metals }
\paragraph{}We begin by deriving a new expression for the dc thermal conductivity $\kappa$ in holographic models of strange metals. Specifically, we consider a general family of holographic Q-lattice models that allow us to study the effects of momentum relaxation whilst retaining a homogeneous bulk metric \cite{Donos:2014uba,Gouteraux:2014hca,Donos:2013eha,Andrade:2013gsa,Davison:2013jba,Vegh:2013sk,Davison:2014lua,Davison:2015bea,Blake:2015epa,Blake:2015hxa}. We will demonstrate that $\kappa$ in these theories depends only on the metric near the horizon: it does not directly depend on the profiles of the matter fields in the black hole solution. This will allow us to establish a relationship near infra red fixed points between the thermal diffusivity $D_T$ and the chaos parameters that is independent of the charge density, magnetic field, or lattice strength.
\paragraph{}We consider the following Q-lattice action, which consists of Einstein-Maxwell-Dilaton gravity coupled to `axion' fields $\chi_i$ that are used to break translational symmetry
\begin{equation}
S=\int d^{d+2}x\,\sqrt{-g}\,\left(R-\frac{1}{2}(\partial\varphi)^{2}-V(\varphi)-\frac{1}{2}W (\varphi)\,(\partial\chi_{ i})^{2}-\frac{1}{4}Z(\varphi)\,F^{2} \right), 
\label{eq:Qaction}
\end{equation}
where ${i} = 1,\ldots,d$ runs over the spatial dimensions of the boundary theory. This action admits homogeneous and isotropic solutions of the form 
\begin{eqnarray}
ds^{2}_{d+2} &=&-f(r)\,dt^{2}+\frac{dr^2}{f(r)} +h(r) dx_{i}^2, \nonumber \\
A=a(r) &dt,& \;\;\;\;\;\; \chi_{i} = k x_{i}, \;\;\;\;\;\;\;\;\; \varphi = \varphi(r), 
\label{eq:Qans}
\end{eqnarray}
where we will assume $f(r_0) = 0 $ corresponds to a black hole horizon with temperature $ 4 \pi T = f'(r_0)$ .
\paragraph{}  The bulk Maxwell field is dual to a conserved $U(1)$ current on the boundary. Integrating the Maxwell equation gives
\begin{equation}
a^{\prime}(r)=\frac{1}{Z(\varphi(r)) h(r)^{d/2} } \rho,
\end{equation}
with $\rho$ the field theory charge density. The axion fields (with $k\ne0$) break translational symmetry of the boundary QFT and leads to momentum relaxation. This results in a finite dc thermoelectric conductivity matrix, whose elements are determined by the solution at the black hole horizon \cite{Blake:2013bqa,Blake:2013owa,Donos:2014uba,Donos:2014cya,Banks:2015wha, Donos:2015gia}
\begin{eqnarray} 
\sigma &=& h(r_0)^{d/2-1} Z(\varphi(r_0))+ \frac{4 \pi \rho^2}{k^2 W(\varphi(r_0)) s}, \nonumber  \\
\alpha &=& \frac{4 \pi \rho}{k^2 W(\varphi(r_0))}, \nonumber  \\
 \bar{\kappa} &=& \frac{4 \pi s T}{k^2 W(\varphi(r_0))},
 \label{thermoresponsemaintext}
\end{eqnarray} 
where $ s = 4 \pi h(r_0)^{d/2}$ is the entropy density.
\paragraph{} These dc conductivities are sensitive to both the metric and the profile and couplings of the matter fields. However the key observation of this paper is that $\kappa$, the thermal conductivity in the absence of electrical current flow, depends only on the background metric. Naively, $\kappa$ depends on the matter fields through
\begin{equation}
{\kappa}\equiv\bar{\kappa}-\frac{T\alpha^2}{\sigma}=\frac{4 \pi s T  Z(\varphi(r_0)) h(r_0)^{d-1}}{\rho^2 + k^2 W(\varphi(r_0)) Z(\varphi(r_0)) h(r_0)^{d-1}}.
\label{kappab}
\end{equation}
But this dependence can be removed by using the equations of motion for the background geometry. In particular, the Einstein equations imply  
\begin{equation}
h^{2 -d/2} (f^{\prime} h^{d/2 - 1})'  -k^{2}W- h Za^{\prime}{}^{2} - f h''  - \frac{{d-2}}{2} f h^{-1}h^{\prime}{}^{2}=0. \\
\label{backgroundeinstein}
\end{equation} 
By evaluating this on the black hole horizon, we can simplify the expression for the thermal conductivity to
\begin{equation}
{\kappa} = 4 \pi \frac{ f' h^{d - 2} }{ (f' h^{d/2 - 1})'}\bigg |_{r_0}, 
\label{kappa}
\end{equation}
which explicitly depends only on the metric. 
Note that the reason this is possible is that the only way the matter fields appear in the thermal conductivity (\ref{kappab}) is through components of the stress tensor, and hence we are always able to eliminate them in favour of the geometry using the Einstein equations. This should be contrasted with the behaviour of the electrical conductivity, which explicitly depends on the matter fields in a way that cannot be re-expressed in terms of the metric.
\section{Thermal Diffusivity and Chaos at Fixed Points}
\label{sec:diffusivity}
\paragraph{} Now that we have established our formula (\ref{kappa}) for the thermal conductivity, it is possible to show that that near generic infra-red fixed points the thermal diffusivity $D_T = \kappa/c_{\rho}$ will be universally related to the chaos parameters. The key point is that (\ref{kappa}) tells us that $\kappa$ is determined solely by the geometry near the infra-red horizon.  Near an infra-red fixed point the entropy density will typically be a power law in temperature. In this case it can also be extracted from the horizon as $c_{\rho} = T (\partial {s} / \partial T)_{\rho} \sim 4 \pi h(r_0)^{d/2}$, with a constant of proportionality dependent upon the scaling exponents characterising the infra red fixed point.  In such cases, it is therefore possible to express $D_T$ entirely in terms of the near-horizon geometry. 
\paragraph{} Similarly, the chaos parameters are also infra-red quantities whose leading small temperature behaviour is determined by the fixed point geometry. Specifically, the chaos parameters of any holographic geometry can be extracted from studying a shock-wave propagating on the black hole horizon \cite{Shenker:2013pqa,Roberts:2014isa,Roberts:2016wdl,Blake:2016wvh}. The Lyapunov time (the inverse of the Lyapunov exponent) is universally given by the temperature of the black hole solution through $\tau_L = (2 \pi T)^{-1}$. The butterfly velocity is model dependent, and can be expressed in terms of the metric near the horizon as
\begin{equation}
v_B^2 = \frac{4 \pi T}{d h'(r_0)}.
\label{vbsquare}
\end{equation}
We therefore conclude that in the infra-red (low temperature) limit the thermal diffusivity and butterfly velocity are related by 
\begin{eqnarray}
D_{T} &\sim& \frac{  f' h^{d/2-1} } {(f'  h^{d/2 - 1} )'} \frac{h'}{h} \bigg|_{r_0} {v_B^2} \tau_L. 
\label{eqnresult}
\end{eqnarray}
As we will show shortly, the power law form of the metric functions $f$ and $h$ near an infra-red fixed point means that the coefficient relating $D_T$ and $v_B^2\tau_L$ in (\ref{eqnresult}) is just a pure number that is independent of all energy scales in the theory (such as the charge density or lattice strength at the fixed point).

\subsection*{Generic fixed points} 

\paragraph{} More concretely, we can consider solutions (\ref{eq:Qans}) that are asymptotically AdS and that flow to Lifshitz and/or hyperscaling-violating geometries in the infra-red. These solutions can be obtained using our action (\ref{eq:Qaction}) by choosing exponential profiles for the potentials
\begin{equation}
 V(\varphi) = - V_0 e^{\delta \varphi} \;\;\;\;\;\; W(\varphi) = W_0e^{\lambda \varphi} \;\;\;\;\;\;\;\ Z(\varphi) = Z_0^2 e^{\gamma \varphi},
 \label{potentials}
\end{equation}
which support solutions of the form \cite{Donos:2014uba,Gouteraux:2014hca,Gouteraux:2011ce,Charmousis:2010zz,Gouteraux:2012yr,Gouteraux:2013oca}
\begin{equation}
f(r) = L_t^{-2} r^{u_1} \bigg( 1 - \frac{r_0^{\Delta}}{r^{\Delta}}\bigg),  \;\;\; h(r) =L_x^{-2} r^{2 v_1}, \;\;\;\ \varphi(r) = {\varphi}_1 \mathrm{log} r, \;\;\; a(r) = A_0 r^{a_1}, \;\;\; \chi_{i} =  k x_{i},
\label{hyperscalingmetric}
\end{equation}
where $\Delta = d v_1 + u_1 - 1$. The space of solutions can be parameterised using a dynamic critical exponent, $z$, and a hyper scaling violating exponent, $\theta$ \cite{Hartnoll:2009ns, Dong:2012se,Huijse:2011ef,Gouteraux:2011ce}. These are related to the power laws in the metric by
\begin{equation}
u_1 = \frac{2 z - 2 \theta/d}{z - 2 \theta/d}, \;\;\;\;\;\;\;\;  2 v_1 = \frac{2 - 2 \theta/d}{z - 2 \theta/d} \;\;\;\;\;\;\;\; \varphi_1^2 = \frac{c}{\left(z - 2 \theta/d\right)^2}
\end{equation}
with $c = (d - \theta) ( 2 z - 2 - 2 \theta/d)$, and $(z, \theta)$ are determined by the exponents in the potentials  (\ref{potentials}). There are four different classes of solution, characterized by whether the charge density or lattice fields are irrelevant or marginal deformations of the infra-red geometry \cite{Gouteraux:2014hca}. When at least one of these deformations is marginal, Lorentz symmetry is broken in the infra-red and so we have $z\ne1$. We will assume this is the case for the remainder of this subsection. 
\paragraph{} For a suitable choice of the parameters $L_t$ and $L_x$ then these Lifshitz/hyperscaling-violating metrics solve the equations of motion of (\ref{eq:Qaction}). The expressions for these parameters take a rather complicated form \cite{Gouteraux:2014hca}, and together with the horizon radius $r_0$ they encode the dependence of the geometry on the dilaton potentials, as well as the values of the charge density and lattice fields. However the full details of these solutions are not necessary for our purposes. Indeed the key point is that whilst the parameters $L_t, L_x, r_0$ will determine the absolute value of the diffusion constant, they do not affect its relationship to $v_B$ and $\tau_L$. 
\paragraph{} To see that this is the case, we can extract the thermal conductivity of our solutions using the formula (\ref{kappa}) and the fixed point geometry (\ref{hyperscalingmetric}) as
\begin{equation}
\kappa = \frac{(z - 2 \theta/d)}{2 z - 2} 4 \pi r_0 h(r_0)^{d/2 - 1}.
\end{equation}
For these solutions the entropy density scales as $s \sim T^{(d-\theta)/z}$ and so we can extract the specific heat as
\begin{equation}
c_{\rho} = \frac{d-\theta}{z} 4 \pi h(r_0)^{d/2} 
\end{equation}
The thermal diffusivity is then
\begin{equation}
D_{T} = \frac{z(z - 2  \theta/d)}{(d - \theta)(2 z-2)} L_x^2 r_0^{1 - 2 v_1}.
\end{equation}
Likewise we can extract the butterfly velocity from (\ref{vbsquare}) as
\begin{equation}
v_B^2 \tau_L = \frac{z - 2\theta/d}{d - \theta}  L_x^2 r_0^{1 - 2 v_1}, 
\end{equation}
from which we see that the relationship
\begin{equation}
D_{T} = \frac{z}{2 z - 2} v_B^2 \tau_L,
\label{kapparesultz}
\end{equation}%
holds independently of any of the details of the bulk solution. As claimed the coefficient of proportionality is simply a pure number, determined only by the dynamical critical exponent of the fixed point.
\paragraph{} It was shown in \cite{Blake:2016sud} that the relationship (\ref{kapparesultz}) held for particle-hole symmetric lattice solutions that flowed to Lifshitz/hyperscaling-violating fixed points in $d = 2$. In fact, we have demonstrated that this result is far more widely applicable. In particular the same relationship holds in finite density solutions, and is completely independent of the marginal deformations describing the charge density and lattice fields at the fixed point. We will shortly see that it continues to hold even in the presence of a magnetic field.
\paragraph{} It is worth contrasting the robustness of this connection between the thermal diffusivity and chaos with attempts to generalise the initial charge diffusion results of \cite{Blake:2016wvh}.  Whilst it was possible to relate the charge diffusion constant of certain particle-hole symmetric theories to chaos, there does not appear to be a simple relationship in a general finite density setting \cite{Blake:2016jnn,Davison:2016ngz,Baggioli:2016pia,Kim:2017dgz,Baggioli:2017ojd}. The thermal diffusivity however can always be expressed in terms of the geometry, and so changing the matter field profiles does not affect our result (\ref{kapparesultz}).
\subsection*{Fixed points with $z=1$} 

\paragraph{} Whilst the above argument is valid at generic fixed points, there are a couple of special cases that require a more careful treatment. The first is that we assumed that either the charge density or lattice fields were marginal, so that our fixed point had $z \neq 1$. For theories where both the charge density and lattice fields are irrelevant deformations it is also possible to construct geometries with $z=1$, for which our result (\ref{kapparesultz}) is ill-defined. This is because $(f' h^{d/2 - 1})^{\prime}|_{r_0}$ vanishes at these fixed points. In this case, the leading low temperature behaviour of $\kappa$ is then determined by the leading irrelevant deformations around the fixed point. In contrast to this, the chaos parameters are still set by the fixed point geometry, and hence there will no longer be a simple connection to $D_T$. Indeed since $D_T$ is now controlled by an irrelevant deformation, it will be parametrically larger than $v_B^2 \tau_L$.\footnote{This was previously found for incoherent charge diffusion in translationally invariant theories \cite{Davison:2017}.}
\subsection*{AdS$_2\times R^{d}$ fixed points}

\paragraph{} A second special case is provided by geometries that flow towards AdS$_2 \times R^{d}$ fixed points in the infra-red. Such geometries can be supported either by the lattice fields or the charge density, and arise if the potentials in (\ref{eq:Qaction}) allow for solutions with a constant scalar $\varphi(r) = \varphi_0$. In this case, $\kappa$ remains finite at the fixed point. However for these geometries the spatial metric is just a constant (i.e. $v_1 =  0$) and so neither the specific heat nor the chaos parameters can be extracted from the fixed point solution. 
\paragraph{} In order to calculate the diffusivity, it is therefore necessary to include the leading irrelevant deformations of the geometry. This analysis was performed in \cite{Blake:2016jnn} where it was found that both $v_B$ and $c_{\rho}$ are determined by the same irrelevant deformation of AdS$_2 \times R^{d}$. As a result it was possible to show that they are always related by 
\begin{equation}
D_{T} = E {v_B^2} \tau_L,
\label{ads2}
\end{equation}
with a coefficient $1/2 < E  \leq 1$ that depends only on the dimension of the leading irrelevant mode. In particular when the leading deformation is a dilatonic mode one finds $E = 1$, which matches the relationship seen in extended SYK models \cite{Gu:2016oyy, Davison:2016ngz}. %
\section{Diffusion in a Magnetic Field} 
\label{sec:magfield}
\paragraph{} We will now generalise the result (\ref{kapparesultz}) to include systems in which time reversal symmetry is broken by an external magnetic field $B$. We note that the holographic approach cannot describe Landau quantization in a magnetic field, and so we assume that $B$ is not so large that such effects are important \cite{Muller:2008qx,Denef:2009yy}. 

We consider the simplest case of 2+1-dimensions in which isotropy is preserved. Each element of the thermoelectric conductivity matrix is therefore now a spatial matrix that can be decomposed into longitudinal and Hall components e.g.~ $\hat{\sigma}_{ij}=\sigma_L\delta_{ij}+\sigma_H\epsilon_{ij}$. The appropriate magneto-hydrodynamic theory of transport is described in Appendix \ref{sec:hydroappendix}.
\paragraph{} In this hydrodynamic theory we find that it is only the longitudinal components of the thermoelectric conductivities that enter in the Einstein relations that define the diffusivity matrix (the Hall conductivities necessarily drop out of these equations on symmetry grounds). The generalisation of the diffusivity (\ref{thermdiffintro}) for theories with a magnetic field is therefore
\begin{equation}
D_T=\frac{\tilde{\kappa}_L}{c_{\rho}},
\label{diffB}
\end{equation}
where $c_{\rho} = T ({\partial s}/{\partial T})_{\rho, B}$ is the specific heat at fixed charge and magnetic field, and the relevant dc thermal conductivity $\tilde{\kappa}_L$ is
\begin{eqnarray} 
 \tilde{\kappa}_L \equiv \bar{\kappa}_L - \frac{T \alpha_L^2}{\sigma_L}.
 \label{kappaB}
\end{eqnarray}
Note that  $\tilde{\kappa}_L$ is \textit{not} simply the longitudinal component of the thermal conductivity matrix $\hat{\kappa} = \hat{\bar{\kappa}} - T \hat{\alpha}{\hat{\sigma}}^{-1} \hat{\alpha}$, but rather is defined by (\ref{kappaBappendix}).
\paragraph{} To study this diffusivity in our holographic Q-lattice models, we first need to generalize our metric ansatz to allow dyonic solutions 
\begin{eqnarray}
ds^{2}_{4} &=&-f(r)\,dt^{2}+\frac{dr^2}{f(r)} +h(r) (dx^2 + dy^2),  \nonumber \\
A=a(r) &dt& + B x dy, \;\;\;\;\;\; \chi_1= k x, \;\;\;\;\;\;  \chi_2 = k y, \;\;\;\;\;\;\;\;\; \varphi = \varphi(r). 
\label{eq:Qansappendix}
\end{eqnarray}
For these solutions, it is still possible to obtain analytic expressions that relate the magnetothermoelectric transport coefficients to the geometry and matter fields on the horizon \cite{Blake:2014yla,Blake:2015ina,Donos:2015bxe,Amoretti:2015gna,Kim:2015wba}. The full expressions now take a rather complicated form, however the final result for the thermal conductivity $\tilde{\kappa}_L$ simplifies to give
\begin{eqnarray}
\tilde{\kappa}_{L} &=& \frac{4 \pi s T Z(\varphi(r_0)) h(r_0)}{\rho^2 + B^2 Z(\varphi(r_0))^2 + k^2 W(\varphi(r_0)) Z(\varphi(r_0)) h(r_0)} .
\end{eqnarray}
Using the bulk equations of motion
\begin{equation}
h f^{\prime \prime}  -k^{2}W- h Za^{\prime}{}^{2} -  Z B^2 h^{-1} - f h''  = 0. \\
\end{equation} 
we find that this can be expressed solely in terms of the background geometry 
\begin{equation}
\tilde{\kappa}_L = \frac{4 \pi f'(r_0)}{f''(r_0)} ,
\label{kappametricmag}
\end{equation}
in exactly the same form as before. Once again the way in which $B$ affects the thermal conductivity is entirely captured by its backreaction on the metric through (\ref{kappametricmag}).  Since the thermal diffusivity takes precisely the same form, then it is clear that our analysis of infra-red fixed points in Section~\ref{sec:diffusivity} will immediately extend to geometries with an external magnetic field.

\paragraph{} In particular, Lifshitz and/or hyperscaling-violating fixed point solutions with a magnetic field have previously been constructed in \cite{Amoretti:2016cad, Kundu:2012jn,Goldstein:2010aw}. The only difference to our previous discussion is that, in addition to the charge and lattice fields, it is now possible for the magnetic field to be a marginal deformation of the fixed point.\footnote{In addition to the solutions discussed in \cite{Amoretti:2016cad, Kundu:2012jn,Goldstein:2010aw}, we also find that when $\theta = 4$ it is possible to construct solutions where the charge density, magnetic field and axions are all marginal deformations.} In this case, the parameters $L_t$ and $L_x$ will also depend on the magnetic field. However, as we have seen, whilst these parameters affect the absolute value of the diffusion constant, they do not change the relationship to the chaos exponents. We therefore have that the relationship (\ref{kapparesultz}) still holds in these solutions, and is independent of the values of the magnetic field, lattice fields or charge density at the fixed point. Similarly the analysis of \cite{Blake:2016jnn} straightforwardly extends to dyonic AdS$_2 \times R^{d}$ solutions and we have that the relationship (\ref{ads2}) also applies for these geometries. 
\section{Discussion}
\paragraph{} In this paper we have studied thermal transport in a general family of homogeneous holographic models of strange metals. We obtained a new expression (\ref{kappa}) for the thermal conductivity of these theories that depended only on the metric near the horizon. This allowed us to show that for generic infra-red fixed points the thermal diffusivity $D_T = \kappa/c_{\rho}$ is related to the chaos exponents through $D_T \sim v_B^2\tau_L$. The coefficient is completely independent of the charge density, lattice strength and magnetic field at the fixed point. Indeed the only exception we found was fixed points with $z=1$, for which the thermal conductivity (\ref{eqnresult}) is ill-defined in the fixed point geometry and hence is sensitive to irrelevant deformations. 
\paragraph{} The remarkable robustness of this result is reminiscent of how the universality of the ratio of the shear viscosity to the entropy density, $\eta/s$, arises in holographic theories \cite{Kovtun:2004de,Iqbal:2008by}. In both cases, there is a transport coefficient which depends only on the infra-red metric and not explicitly on the matter fields. When expressed in terms of an appropriate thermodynamic quantity, one then finds very simple expressions for these transport coefficients. Unlike the universality of $\eta/s$, our result applies only at low temperatures near infra-red fixed points. However, in other ways our result is more general.  In particular, it does not rely on the state being translationally invariant and we demonstrate in Appendix~\ref{anisotropicappendix} that it continues to hold in anisotropic theories. Both of these are situations in which the original $\eta/s$ result can be badly violated \cite{Rebhan:2011vd,Mamo:2012sy, Jain:2015txa,Jain:2014vka,Davison:2014lua,Hartnoll:2016tri,Alberte:2016xja, Burikham:2016roo}. Additionally, we show in Appendix~{\ref{dbiappendix}} that our expression (\ref{kappa0}) also holds for DBI Q-lattice solutions, and so our conclusions are not dependent on the choice of Maxwell action for the gauge field. 
\paragraph{} We noted that for finite density theories the Einstein relation  (\ref{thermdiffintro}) for our thermal diffusivity does not a priori correspond to an eigenvalue of the thermoelectric diffusivity matrix. However for the holographic models we have studied in this paper, $D_T$ is equal to one of the eigenvalues of the diffusion matrix near the infra-red fixed point (see Appendix \ref{sec:hydroappendix}). We expect that the connection between chaos and an eigenvalue of the diffusivity matrix in an `incoherent' limit found in \cite{Baggioli:2016pia,Kim:2017dgz,Blake:2016jnn,Baggioli:2017ojd} can also be understood as consequences of a simple equation for $\kappa$, as we have outlined.
\paragraph{} In future work it would be very interesting to determine to what extent our results can be generalised to yet more complicated systems. Two of the most promising avenues to pursue are to investigate inhomogeneous solutions  \cite{Horowitz:2012ky,Donos:2014yya} and higher derivative theories of gravity \cite{Donos:2017oym}. Inhomogeneous holographic theories  and SYK chains  have been studied in \cite{Lucas:2016yfl, Gu:2017ohj}, and the relation between a `disorder-averaged' diffusivity and butterfly velocity found to depend on the profile of inhomogeneities. However it would be interesting to understand if a relationship like (\ref{eq:firsteq}) still holds in terms of a local diffusivity and butterfly velocity. A natural starting point for this would be to determine if and how our observation that the thermal conductivity can be expressed in terms of the metric generalises to inhomogeneous cases.
\paragraph{} Finally, as we mentioned in our introduction, similar connections between the thermal diffusivity and chaos have also been observed in non-holographic models. For instance critical Fermi surface models and extended SYK models both have a thermal diffusivity that is given by $v_B^2 \tau_L$, up to an order one coefficient \cite{Patel:2016wdy,Jian:2017unn,Blake:2016jnn,Narayan:2017qtw,Davison:2016ngz,Gu:2016oyy} (see \cite{Bohrdt:2016vhv,Patel:2017vfp,Chowdhury:2017jzb,Leviatan:2017vur} for related work in other systems). Note although the coefficient in (\ref{kapparesultz}) is very simple, this expression is not expected to be universal across all quantum field theories with the same dynamical critical exponent. In particular the models studied in \cite{Patel:2016wdy} have $z=3/2$ but with a diffusivity $D_T = 0.42 v_B^2 \tau_L$. Nevertheless, the essential point is that, as in our holographic examples, the coefficient relating the diffusivity to the chaos exponents is independent of the UV parameters of the system. It would be fascinating to develop a more complete understanding of why this is the case, and to see if these ideas can be applied to experimental systems \cite{Bruin2013} such as underdoped YBCO whose thermal diffusivity was recently reported in \cite{Zhang:2016ofh}.
\acknowledgments{We are grateful to Simon Gentle and Blaise Gouteraux for helpful discussions. RD is supported by the Gordon and Betty Moore Foundation EPiQS Initiative through Grant GBMF\#4306. SS and RD are supported by MURI grant W911NF-14-1-0003 from ARO. The work of MB is supported by the Office of High Energy Physics of U.S. Department of Energy under grant Contract Number  DE-SC0012567.  Research at Perimeter Institute is supported by the Government of Canada through Industry Canada and by the Province of Ontario through the Ministry of Research and Innovation. SS also acknowledges support from Cenovus Energy at Perimeter Institute. }

\appendix

\section{Diffusive Processes in a Metal}
\label{sec:hydroappendix}

\paragraph{}In a strongly interacting system with momentum relaxation (i.e.~one without translational symmetry) the only long-lived modes are long wavelength fluctuations in the charge $\delta \rho$ and energy density $\delta \varepsilon$. There is a very simple effective theory for the dynamics of these modes  \cite{Hartnoll:2014lpa} -- on long distances and timescales they are just described by a pair of coupled diffusion equations for these conserved charges. 
\paragraph{} To obtain these equations, then it is convenient to first change variables and study perturbations of heat $\delta s$ rather than energy using 
\begin{equation}
T\delta s\equiv\delta\varepsilon-\mu\delta\rho.
\end{equation}
The effective theory is then given by the conservation equations
\begin{equation}
\partial_t\delta\rho+\nabla\cdot j=0,\;\;\;\;\;\;\;\;\;\;\partial_t\delta s+\frac{1}{T}\nabla\cdot j_Q=0,
\label{conservationlaws}
\end{equation}
together with the constitutive relations
\begin{equation}
\label{eq:constitutiverelations}
j=-\sigma\left(\nabla\mu- E\right)-\alpha\nabla T,\;\;\;\;\;\;\;\;\;\; j_Q=-\alpha T\left(\nabla\mu- E\right)-\bar{\kappa}\nabla T, 
\end{equation}
where $j$ and $j_Q$ are the charge and heat currents, $\sigma,\alpha$ and $\bar{\kappa}$ are the DC thermoelectric conductivities, and $E$ is an external electric field.  
Note that this is qualitatively different to the effective theory of translationally invariant theories, for which there is an additional long-lived mode corresponding to long wavelength perturbations of the momentum density that must also be included. 
\paragraph{}In the absence of an external field one then finds a pair of coupled diffusion equations 
\begin{equation}
\begin{aligned}
\partial _t\left( \begin{array}{c} \delta\rho  \\ \delta s \end{array} \right)=D\cdot\nabla^2\left( \begin{array}{c} \delta\rho  \\ \delta s  \end{array} \right),
\end{aligned}
\end{equation}
where the diffusivity matrix $D$ is given by the matrix product of the thermoelectric conductivity matrix $\Sigma$ and the inverse of the thermodynamic susceptibility matrix $\chi_{s}$
\begin{equation}
D=\Sigma\cdot\chi_{s}^{-1}=\left( \begin{array}{cc} \sigma & \alpha \\ \alpha & \bar{\kappa}/T \end{array} \right)\cdot\left( \begin{array}{cc}  \chi & \xi \\ \xi & c_{\mu}/T \end{array} \right)^{-1},
\label{diffmatrix}
\end{equation}
with
\begin{equation}
\chi = \bigg( \frac{\partial \rho }{\partial \mu} \bigg)_{T},  \;\;\;\;\;\;\;\;\; \xi = \bigg( \frac{\partial s}{\partial \mu} \bigg)_{T}, \;\;\;\;\;\; c_{\mu} = T  \bigg( \frac{\partial s}{\partial T} \bigg)_{\mu}.
\end{equation}
\paragraph{} 
These diffusion equations can be decoupled using eigenmodes of $D$, that describe linear combinations of these perturbations that diffuse independently. The two thermoelectric diffusivities, $D_+,D_-$ ( the eigenvalues of $D$ ) now satisfy  \cite{Hartnoll:2014lpa}
\begin{equation}
\begin{aligned}
D_++D_-&=\frac{\kappa}{c_\rho}+\frac{\sigma}{\chi}+\frac{T\sigma}{c_{
\rho}}\left(\frac{\alpha}{\sigma}-\left(\frac{\partial s}{\partial\rho}\right)_T\right)^2,\\
D_+D_-&=\frac{\kappa}{c_\rho}\frac{\sigma}{\chi},
\end{aligned}
\label{eigenvalues}
\end{equation}
where $\kappa$ is the open circuit dc thermal conductivity
\begin{equation}
\kappa\equiv-\frac{j_Q}{\nabla T}\Bigr|_{j=0}=\bar{\kappa}-\frac{T \alpha^2}{\sigma},
\end{equation} 
and $c_\rho$ is the heat capacity at constant density
\begin{equation}
c_\rho=T\left(\frac{\partial s}{\partial T}\right)_\rho = c_{\mu} -  \frac{T \xi^2}{\chi}.
\end{equation}
\paragraph{} The thermal diffusivity $D_T = \kappa/c_{\rho}$ we have calculated in this paper is then only equivalent to an eigenvalue of the diffusion matrix when the thermoelectric cross terms in (\ref{eigenvalues}) can be neglected. This will certainly be the case provided
\begin{equation}
\chi^{-1} \gg \frac{T}{c_{\rho}}\left(\frac{\alpha}{\sigma}-\left(\frac{\partial s}{\partial\rho}\right)_T\right)^2\\
\label{decouplingcondition}
\end{equation}
If this condition is satisfied then the $\sigma/\chi$ term dominates over the mixing terms in (\ref{eigenvalues}) and hence we will have that the two eigenmodes are simply given by $D_c = \sigma/\chi$ and $D_T = \kappa/c_{\rho}$. We will shortly demonstrate that this condition is always satisfied in the infra-red limit of the fixed points we studied in section \ref{sec:diffusivity}. In the low temperature limit of these models, $\kappa/c_{\rho}$ then indeed coincides with an eigenvalue of the diffusivity matrix.
\paragraph{}Note that in general this $\kappa/c_{\rho}$ piece of the diffusivity matrix can be directly extracted by turning on an external electric field that enforces the condition that the charge density is static i.e. $\partial_{t} \delta \rho = 0$. From equation (\ref{eq:constitutiverelations}), this can be achieved with the choice 
\begin{equation}
E=\nabla\mu+\frac{\alpha}{\sigma}\nabla T.
\end{equation}
With this constraint satisfied, the diffusion equations then reduce to the condition that temperature fluctuations 
$
\delta T\equiv\left(\partial T/\partial s\right)_\rho\delta s+\left(\partial T/\partial\rho\right)_s\delta\rho,
$
obey the diffusion equation
\begin{equation}
\partial_t\delta T=D_T\nabla^2\delta T, \;\;\;\;\;\;\;\; D_T = \frac{\kappa}{c_{\rho}}.
\end{equation}
\subsection*{Diffusion in a Magnetic Field} 
\paragraph{} For 2+1 dimensional theories, it is straightforward to extend our discussion of diffusion to systems with an external magnetic field. The conservation laws (\ref{conservationlaws}) remain valid, but we must modify our constitutive relations to include the off-diagonal elements of the conductivity tensors
\begin{equation}
\label{eq:constitutiverelationsB}
j_i=-\hat{\sigma}_{i j} \left(\nabla_j \mu-E_j\right)-\hat{\alpha}_{i j} \nabla_j T,\;\;\;\;\;\;\;\;\;\; j_{Q_{i}}=-\hat{\alpha}_{ij} T\left(\nabla_j\mu-E_j\right)-\hat{\bar{\kappa}}_{ij}\nabla_j T, 
\end{equation}
where $i,j$ run over the two spatial directions. The thermoelectric conductivities in (\ref{eq:constitutiverelationsB}) are now matrices that (for isotropic theories) can be decomposed into longitudinal and Hall components e.g. $\hat{\sigma}_{ij}=\sigma_L\delta_{ij}+\sigma_H\epsilon_{ij}$. 
\paragraph{} In the absence of the electric field, one now finds that the fluctuations are described by the diffusion matrix
\begin{equation}
D=\Sigma\cdot\chi_{s}^{-1}=\left( \begin{array}{cc} \sigma_L & \alpha_L \\ \alpha_L & \bar{\kappa}_L/T \end{array} \right)\cdot\left( \begin{array}{cc}  \chi & \xi \\ \xi & c_{\mu}/T \end{array} \right)^{-1},
\label{diffmatrixB}
\end{equation}
where the magnetic field should be held fixed when taking thermodynamic derivatives.  Note that the Hall conductivities completely drop out of the diffusivity matrix. The diffusion equations therefore take the same form as without the magnetic field, provided one replaces the usual thermoelectric conductivities $\sigma, \alpha, \bar{\kappa}$ with their longitudinal components $\sigma_L$, $\alpha_L$, $\bar{\kappa}_L$.  
\paragraph{} In particular the eigenvalues of $D$ now satisfy
\begin{equation}
\begin{aligned}
D_++D_-&=\frac{\tilde{\kappa}_L}{c_\rho}+\frac{\sigma_L}{\chi}+\frac{T\sigma}{c_{
\rho}} \left(\frac{\alpha_L}{\sigma_L}-\left(\frac{\partial s}{\partial\rho}\right)_T\right)^2,\\
D_+D_-&=\frac{\tilde{\kappa}_L}{c_\rho}\frac{\sigma_L}{\chi},
\end{aligned}
\label{eigenvaluesB}
\end{equation}
where the `longitudinal' thermal conductivity in these expressions
\begin{eqnarray} 
 \tilde{\kappa}_L = -\frac{j_{Q_x}}{\nabla_x T} \bigg|_{\nabla_{y} T, j_x =0, E_y=0} = \bar{\kappa}_L - \frac{T \alpha_L^2}{\sigma_L},
 \label{kappaBappendix}
\end{eqnarray}
is the generalization of $\kappa$ that appears in the diffusion equations. As we noted in the main text, this is distinct from first computing the open-circuit thermal conductivity $\hat{\kappa} = \hat{\bar{\kappa}} - T \hat{\alpha}{\hat{\sigma}}^{-1} \hat{\alpha}$ and then taking the longitudinal part. 
\paragraph{} Once again for fixed points with a magnetic field we find that the thermoelectric mixing terms in (\ref{eigenvaluesB}) are subleading at low temperatures. The eigenvalues of the diffusivity matrix are therefore now given by $\sigma_L/\chi$ and $\tilde{\kappa}_L/c_{\rho}$ in the infra-red limit. To directly extract the $\tilde{\kappa}_L/c_{\rho}$ component of the diffusivity matrix, we can again consider turning on an external electric field such that $\partial_{t} \delta \rho = 0$. Under the extra condition that this electric field has vanishing curl, then the diffusion equations now reduce to
\begin{equation}
\partial_t\delta T=D_T\nabla^2\delta T, \;\;\;\;\;\;\;\; D_T = \frac{\tilde{\kappa}_L}{c_{\rho}}.
\end{equation}
\subsection*{Thermoelectric mixing terms}
\paragraph{} We now wish to justify that in the low-temperature limit of our finite density fixed points we always satisfy the condition (\ref{decouplingcondition}) and hence the thermal diffusivity $D_T = \kappa/c_{\rho}$ corresponds to an eigenvalue of the diffusivity matrix. To see this we note that all the quantities in the thermoelectric mixing terms are infra-red quantities  whose leading temperature dependence can be extracted from the near-horizon solution (\ref{hyperscalingmetric}). 

\paragraph{} In particular if we assume the charge density is a marginal deformation then we can extract $c_{\rho}$ and $(\partial s/\partial \rho)_T$ by differentiating the entropy density at the fixed point. We then have the scalings $c_{\rho} \sim (\partial s/\partial \rho)_T \sim T^{(d - \theta)/z}. $ Likewise, from the explicit formulae (\ref{thermoresponsemaintext}) for the thermoelectric transport coefficients we deduce that $\alpha/\sigma \sim T^{(d- \theta)/z}$ has the same scaling at generic fixed points. The mixing terms in the Einstein relations are then proportional to 
\begin{equation}
\frac{T}{c_{\rho}} \left(\frac{\alpha}{\sigma}-\left(\frac{\partial s}{\partial\rho}\right)_T\right)^2 \sim T^{(z + d - \theta)/z} 
\label{crosstermscaling}
\end{equation}
which always vanishes in the infra-red limit. Similarly if the charge density is an irrelevant deformation then these mixing terms will be even further suppressed at low temperatures.
 \paragraph{} In contrast, the low temperature behaviour of the charge susceptibility $\chi^{-1}$ is not controlled solely by the IR fixed point, but receives contributions from all parts of the geometry. In particular there will be a temperature independent piece coming from the UV region and so we will have $\chi^{-1} \sim T^{0}$ in these geometries. As such (\ref{decouplingcondition}) is always satisfied at low temperatures for theories that flow to one of these fixed points. Provided we work with the longitudinal conductivities $\sigma_L, \alpha_L$ then this scaling analysis goes through completely unchanged for dyonic solutions with a magnetic field, and so the thermoelectric cross terms can also be neglected at low temperatures in this case.\footnote{The mixing terms can also be neglected for AdS$_2\times R^{d}$ geometries due to the Kelvin formula $\alpha/\sigma = ({\partial s}/{\partial\rho})_T$ satisfied by the thermoelectric conductivities \cite{Blake:2016jnn,Davison:2016ngz}. For dyonic AdS$_2$ geometries we find $\alpha_L/\sigma_L = ({\partial s}/{\partial\rho})_{T,B}$ and so this generalises to theories with a magnetic field.}  

\section{Anisotropic Q-Lattice Models}
\label{anisotropicappendix}
\paragraph{} In this appendix we will show that universal relations between the thermal diffusivities and butterfly velocities also hold in anisotropic Q-lattice solutions. In order to be explicit, we will study theories with two spatial directions $x$ and $y$. We therefore consider  the following generalisation of our Q-lattice action
\begin{equation}
S=\int d^{4}x\,\sqrt{-g}\,\left(R-\frac{1}{2}(\partial\varphi)^{2}-V(\varphi)-\frac{1}{2}W_1 (\varphi)\,(\partial\chi_{ 1})^{2}-\frac{1}{2}W_2 (\varphi)\,(\partial\chi_{ 2})^{2} - \frac{1}{4}Z(\varphi)\,F^{2} \right), \nonumber
\label{eq:Qactionanisotropic}
\end{equation}
and are now interested in homogeneous but anisotropic solutions of the form 
\begin{eqnarray}
ds^{2}_{4} &=&-f(r)\,dt^{2}+\frac{dr^2}{f(r)} +h_1(r) dx^2 + h_2(r) dy^2, \nonumber \\
A=a(r) &dt,& \;\;\;\;\;\; \chi_{1} = k_1 x, \;\;\;\ \chi_{2} = k_2 y, \;\;\;\;\;  \varphi = \varphi(r).
\label{eq:Qanisotropic}
\end{eqnarray}
The temperature of these black holes is given by $4 \pi T = f'(r_0)$. The charge density is $\rho = \sqrt{h_1 h_2} Z a'$ and the entropy density is $s = 4 \pi \sqrt{h_1 h_2}(r_0)$.
\paragraph{} For these theories we can now define two thermal diffusivities
\begin{equation}
D_{xx} = \frac{\kappa_{xx}}{c_{\rho}} \;\;\;\\;\; D_{yy} = \frac{\kappa_{yy}}{c_{\rho}}
\end{equation}
corresponding to diffusion along the $x$ and $y$ directions respectively. The DC conductivities of these geometries can again be related to the metric and matter fields at the black hole horizon. In particular the open-circuit conductivity in the $x$ direction is now given by 
\begin{equation}
{\kappa_{xx} }\equiv \bar{\kappa}_{xx}-\frac{T\alpha_{xx}^2}{\sigma_{xx}}=\frac{4 \pi s T  Z(\varphi(r_0)) h_2(r_0)}{\rho^2 + k_1^2 W_1(\varphi(r_0)) Z(\varphi(r_0)) h_2(r_0)},
\label{kappaanisotropic}
\end{equation}
whilst the conductivity along the $y$ direction follows from swapping the indices 1 and 2 in this expression.
\paragraph{}As we saw in the main text, the key to establishing a general connection between the diffusion constant and chaos was to express this thermal conductivity entirely in terms of the bulk geometry. Using the Einstein equations we find (\ref{kappaanisotropic}) can be written as
\begin{equation}
{\kappa}_{xx} = \frac{4 \pi f' h_2 h_1^{-1}}{ (f' h_2^{1/2} h_1^{-1/2})' }\bigg|_{r_0} ,
\label{kappaanisotropicmetric}
\end{equation}
and so the same is true for these anisotropic theories. It is then straightforward to evaluate (\ref{kappaanisotropicmetric}) for geometries that flow to anisotropic power law solutions in the infra-red. These take the form 
\begin{equation}
f(r) = L_t^{-2} r^{u_1} \bigg( 1 - \frac{r_0^{\Delta}}{r^{\Delta}}\bigg),  \;\;\; h_1(r) =L_x^{-2} r^{2 v_1}, \;\;\;\ h_2(r) =L_y^{-2} r^{2 v_2},\label{anisotropicpowerlawmetric}
\end{equation}
with $\Delta =  v_1 + v_2 + u_1 - 1$. The thermal conductivity of these solutions is just
\begin{equation}
\kappa_{xx} = \frac{1}{u_1 - 2 v_1 } 4 \pi L_x L_y^{-1} r_0^{1 + v_2 - v_1}.
\end{equation}
The Einstein relation $D_{xx} = \kappa_{xx}/c_{\rho}$ then gives the thermal diffusivity
\begin{equation}
D_{xx} = \frac{u_1 - 1}{(v_1 + v_2)(u_1 - 2 v_1) }  L_x^2 {r_0}^{1 - 2 v_1}, 
\label{diffxx}
\end{equation}
whilst the diffusivity in the $y$ direction is 
\begin{equation}
D_{yy} =  \frac{u_1 - 1}{(v_1 + v_2)(u_1 - 2 v_2) } L_y^2 r_0^{1- 2 v_2}. 
\label{diffyy}
\end{equation} 
\subsection*{Butterfly Velocity in Anisotropic Theories} 
\paragraph{} Now that we have these diffusivities, we can compare with the relevant butterfly velocities along the $x$ and $y$ directions.  These can be calculated by considering the equation for a shock wave perturbation $\delta g_{u u}$ on an anisotropic horizon \cite{Blake:2016wvh,Ling:2016ibq} 
\begin{equation}
(h^{i j}(r_0) \partial_{i}\partial_{j}  - m^2 ) \delta g_{uu} (t_w, \vec{x}) \sim  E e^{\frac{2 \pi}{\beta} t_w} \delta(\vec{x}),
\label{poisson}
\end{equation}
where the right hand side describes the stress tensor of an in falling particle of boundary energy density $E$ that sources the shock-wave geometry. Here $h^{ij}(r_0)$ is the inverse-spatial metric on the horizon and the effective mass is given by
\begin{equation}
m^2 = \pi T \bigg( \frac{ h_2 h_1 ' + h_1 h_2 ' }{h_2 h_1} \bigg) \bigg|_{r_0}. 
\label{musquare}
\end{equation} 
This equation implies a Lyapunov time $\tau_L = (2 \pi T)^{-1}$ and the anisotropic butterfly velocities 
\begin{equation}
v_x = \frac{2 \pi T}{\sqrt{h_1} m}\bigg|_{r_0}, \;\;\;\;\;\;\;\;\; v_y = \frac{2 \pi T}{\sqrt{h_2} m }\bigg|_{r_0}.
\end{equation}
For our power law geometries this just leads to  
\begin{equation}
v_x^2 \tau_L = \frac{1}{v_1 + v_2 } L_x^2 r_0^{1 - 2{v_1}}, \;\;\;\;\; v_y^2 \tau_L = \frac{1}{v_1 + v_2 } L_y^2 r_0^{1 - 2{v_2}}.
\label{anisobutterfly}
\end{equation}
And so we establish the relationships
\begin{equation}
D_{xx} = a {v_x^2} \tau_L, \;\;\;\;\;\;\;\;\;\  D_{yy } = b v_y^2 \tau_L, \;\;\;\;\;\;\;\;\;\ a = \frac{u_1 - 1}{u_1 - 2 v_1}, \;\;\;\;\;\;\;\;\;\;\;\;  b = \frac{u_1 - 1 }{u_1 - 2 v_2}.
\end{equation}
At low temperatures we will therefore again have universal relations between these diffusivities and the chaos exponents. Indeed the essential point is that the anisotropic butterfly velocities $v_x, v_y$ precisely account for the different dependence on $r_0, L_x,L_y$ that appears in the diffusivities (\ref{diffxx}) and (\ref{diffyy}). The only difference to the isotropic case is then that the order one coefficients $a,b$ can differ between the $x$ and $y$ directions if we have different power laws in the metric (i.e. $v_1 \neq v_2$).%
\section{DBI Q-Lattice Models}
\label{dbiappendix}

\paragraph{}In this appendix we show that the simple expression (\ref{kappa0}) for $\kappa$ in terms of the metric near the horizon can apply in theories with more general matter actions. Specifically, we consider replacing the Maxwell action for the U(1) gauge field with a DBI-like action
\begin{equation}
\nonumber
S=\int d^4x\sqrt{-g}\left(R-\frac{1}{2}(\partial\varphi)^2-V(\varphi)-\frac{1}{2}W(\varphi)(\partial\chi_i)^2-Z_1(\varphi)\sqrt{-\det(g+Z_2(\varphi)F)}\right),
\end{equation}
and again look for solutions of the form (\ref{eq:Qansappendix}). The charge density is given by
\begin{equation}
\rho=a'Z_1Z_2^2\sqrt{\frac{h^2+B^2Z_2^2}{1-Z_2^2{a'}^2}}.
\end{equation}

\paragraph{} To compute the magneto-thermoelectric conductivities, we follow \cite{Donos:2014cya,Blake:2015ina} and study the linearised fluctuation equations with the ansatz
\begin{equation}
\begin{aligned}
&\delta A_x=\left(-E+\xi a\right)t+\delta a_x(r),\;\;\;\;\;\;\;\;\delta A_y=\delta a_y(r),\\
&\delta g_{xt}=-\xi tf(r)+h(r)\delta h_{xt}(r),\;\;\;\;\;\;\delta g_{yt}=h(r)\delta h_{yt}(r),\\
&\delta g_{r{x^i}}=h(r)\delta h_{r{x^i}}(r),\;\;\;\;\;\;\;\;\;\;\;\;\;\;\;\;\;\;\;\;\;\delta\chi_i=\delta\chi_i(r).
\end{aligned}
\end{equation}
where $E$ and $\xi$ are an electric field and temperature gradient in the $x$-direction. There are two radially conserved quantities $j_x$ and $j_{Qx}$ which we identify with the longitudinal electrical and thermal currents
\begin{equation}
\begin{aligned}
&j_x=-\frac{hZ_1Z_2^2}{\sqrt{\left(h^2+B^2Z_2^2\right)\left(1-Z_2^2{a'}^2\right)}}\left(f\left(\delta a_x'+B\delta h_{yr}\right)+ha'\delta h_{xt}\right),\\
& j_{Qx}=f^2\left(\frac{h}{f}\delta h_{xt}\right)'-aj_x.
\end{aligned}
\end{equation}
We can then impose ingoing boundary conditions at the horizon to obtain the longitudinal conductivities
\begin{equation}
\begin{aligned}
\sigma_L=\frac{m^2 hW\left(\rho^2+B^2Z_1^2Z_2^4+m^2WZ_2X\right)}{h^2m^4W^2+B^2\left(m^4W^2Z_2^2+B^2Z_1^2Z_2^4+\rho^2+2m^2WZ_2X\right)}\Biggr|_{r_0},\\
\alpha_L=\frac{4\pi Wh^2m^2\rho}{h^2m^4W^2+B^2\left(m^4W^2Z_2^2+B^2Z_1^2Z_2^4+\rho^2+2m^2WZ_2X\right)}\Biggr|_{r_0},\\
\bar{\kappa}_L=\frac{16\pi^2hT\left(h^2m^2W+B^2Z_2\left(m^2WZ_2+X\right)\right)}{h^2m^4W^2+B^2\left(m^4W^2Z_2^2+B^2Z_1^2Z_2^4+\rho^2+2m^2WZ_2X\right)}\Biggr|_{r_0},
\end{aligned}
\end{equation}
where $X=\sqrt{\rho^2+\left(h^2+B^2Z_2^2\right)Z_1^2Z_2^2}$. Combining these, we find that
\begin{equation}
\label{eq:DBIkappaL1}
\tilde{\kappa}_L=\frac{16\pi^2hZ_2T\sqrt{\rho^2+\left(h^2+B^2Z_2^2\right)Z_1^2Z_2^2}}{\rho^2+B^2Z_1^2Z_2^4+m^2WZ_2\sqrt{\rho^2+\left(h^2+B^2Z_2^2\right)Z_1^2Z_2^2}}\Biggr|_{r_0}.
\end{equation}

\paragraph{} The Einstein equations for the DBI-like action imply that
\begin{equation}
hf''-m^2W-\frac{Z_1Z_2^2\left(B^2+h^2{a'}^2\right)}{\sqrt{\left(h^2+B^2Z_2^2\right)\left(1-Z_2^2{a'}^2\right)}}-fh''=0.
\end{equation}
Evaluating this on the horizon gives an equation that can be used to simplify the expression (\ref{eq:DBIkappaL1}) for $\tilde{\kappa}_L$ to
\begin{equation}
\tilde{\kappa}_L=4\pi\frac{f'(r_0)}{f''(r_0)}.
\end{equation}
The result that this thermal conductivity can be expressed solely in terms of the near-horizon metric is therefore not special to the matter action we examined in the main text, but applies also to this DBI case. We therefore expect the thermal diffusivity to be related to the chaos parameters near infra-red fixed points of this theory. Some infra-red fixed points of the theory were constructed in \cite{Pal:2012zn,Tarrio:2013tta}.

\bibliographystyle{JHEP}
\bibliography{temperature-diffusivity}

\end{document}